\documentclass[mathleft
]{an}
\usepackage{graphicx}
\usepackage{times}
\newcommand{\dd}{ {\mathrm d} }

\begin{document}
\Pagespan{789}{}
\Yearpublication{2006}%
\Yearsubmission{2005}%
\Month{11}%
\Volume{999}%
\Issue{88}%
\title{Searching Extra Dimensions in Compact Stars}

\author{G. G. Barnaf\"oldi\thanks{Corresponding author:
  \email{bgergely@rmki.kfki.hu}\newline}
\and  P. L\'evai
\and B. Luk\'acs
}
\titlerunning{Searching Extra Dimensions in Compact Stars}
\authorrunning{G.G. Barnaf\"oldi \& P. L\'evai \& B. Luk\'acs}
\institute{
Research Institute for Particle and Nuclear Physics
H-1525 Budapest, P.O.Box 49, Hungary}

\received{20 March 2007}
\accepted{xx March 2007}
\publonline{later}


\abstract{%
The electro-magnetic and particle radiation detected from the direction
of Cygnus X3 raise the question of the existence of special
long-lived, neutral particles. We investigate the origin of these
particles in a special approach: the source object may contain
compactified extra dimension and these particles are messengers
of this state. We describe hyperon stars in $3+1$ dimension and 
introduce the description of a compact stars in  
$3+1_c+1$ dimensional space-time. }

\maketitle
\section{Introduction}

The understanding of the Cygnus X3 is still in a shroud of mystery (Collins 
2000). Although, both electro-magnetic radiations (radio, IR, optical, 
X-ray) and additional particle radiations (muons and neutrinos), 
have been observed since 1981 from the direction of the Cygnus X3 (J2000
$\alpha=308.1^{\circ}$, $\delta=40.96^{\circ}$), however the 
theoretical description of this object and the global understanding of the 
experimental data remained an open question even these days. 

It is commonly accepted, that Cygnus X3 is a galactic, tight binary 
star system (with $\sim 5$ solar radii), which consists of a compact object 
($\sim 3 M_{\odot}$) and a $\sim 15 M_{\odot}$ mass companion Wolf-Rayet 
(WN7 or 8) star with huge mass loss. On the other hand, the Cygnus X3 is 
known as a 'micro-quasar', a small quasar in our galaxy, emitting strong 
radio flares and highly collimated relativistic jets towards Earth 
(Allison et al. 1998; Marshak et al. 1985, 2000).       

Muon shower data have been collected between 1981 and 2002 by 
the deep underground muon detectors (e.g. SOUDAN, NUSEX, and SOUDAN2).
The measured muon flux clearly correlates to the 4.8 hours X-ray cycle caused 
by the eclipsing (or wind) of the compact star by its companion 
(Hillas 1984). Furthermore, eruptions (flares) have been fo\-und  
episodically and their lengths are varying in a wide range, as collected 
in Table \ref{cyg-exp}.

In recent article we concentrate on the primary origin of the muon showers,
if these showers are caused by particles produced in the Cygnus X3 and survived
the trip to Earth. These particle-like objects are named 'cygnets', and
they must have the following properties: 
\begin{enumerate}  
\item[(i)] {\it neutral}, otherwise the path would be curved by the 
ga\-lac\-tic magnetic field which would have deflected their 
arrival directions or have randomized it; 
\item[(ii)] {\it stable} or {\it long-lived}, otherwise they would 
decay along their 10 kpc ($\sim30,000$ ly) long way; 
\item[(iii)] {\it strongly-interacting}, since they produce 
hadron \mbox{showers} including pions, which decay into muons. 
\end{enumerate}

\begin{table}
 \centering
\caption{The main observed radio flares of Cygnus X3, based on the works of 
Allison et al. (1998), Marshak et al. (1985, 2000), Mart\'\i ~et al. (2001),
and Miller-Jones et al. (2004).}
\label{cyg-exp}
\begin{tabular}{llc}
\hline
Start (date) & End (date)& Duration (day)\\
\hline
1989 Jun 1 & 1989 Aug 14 & 75 \\
1990 Aug 11 & 1990 Aug 31 & 21 \\
1990 Oct 1& 1990 Oct  22 & 22 \\
1991 Jan 18 & 1991 Mar 19 & 163  \\
1991 Jun 19 & 1991 Jun 30  & 12\\
1991 Jul 24 & 1991 Aug 29 & 37\\
1992 Sep 1 & 1992 Sep 11 & 11 \\
1994 Feb 20 & 1994 Mar 11 & 20 \\
1997 Feb 1 & 1997 Feb  16 & 16\\
1997 Jun 10 & 1997 Jun  25 & 16 \\
1998 Mar 14 & 1998 Mar  20 & 7 \\
2000 Oct 21 & 2000 Oct 21 & 1\\
2000 Nov 4 & 2000 Nov 5 & 2 \\
2000 Nov 19 & 2000 Nov 20 & 2 \\
2001 Sep 9 & 2001 Sep 26 & 18\\
\hline
\end{tabular}
\end{table}

New particles were suggested to solve the 'Cygnus X3-puzzle': 
the '{\it uds-nuggets}' proposed by Witten (1984); the 
'{\it H$^0$ dibaryons}' listed by Baym (1985), Kondratiuk (1986), and 
BNL (2000). However these particles have not been detected yet 
experimentally. 

In this paper we present another approach and study the appearence of  
microscopical quanta with extra dimensional aspect: these
particles are produced inside the compact star, where usual 
description in $3+1$ dimensional space-time is extended by an extra 
Kaluza\,--\,Klein (KK) like compactified fifth dimension ($1_c$).

\section{Strangeness or Extra Dimension?}

The 'strange-star' aspect of the Cygnus X3 has been introduced and discussed
several times (Grassi 1988; Olinto 1991; Glendenning 1997; Weber 1999): the 
basic idea is to assume a neutron or quark star with an inner core, which 
contains large number of strange quarks. Although, these strange quark stars 
usually become instable in the theoretical calculations from the point of 
view of radial excitations (Glendenning 1997; Priszny\'ak et al. 1994; 
Barnaf\"oldi et al. 2003; Luk\'acs et al. 2003), however quickly rotated 
quark stars may be stabilized. The existence of millisecond pulsars supports this expectation. 

If such a strange quark star has a tight binary companion, which can reach 
the Roche lobe, then matter can flow to the compact object. The particles 
can gain enough kinetic energy and become relativistic, thus they can kick 
out matter below the neutron surface of the strange star. The out-kicked 
relativistic strange matter can fly away and reach our detectors in Earth.
We claim that this mechanism is the most probable to create accelerated 
particle packages with extra strangeness content, which packages reach the 
atmosphere of the Earth time-to-time,
and which packages are responsible for the measured intense muon showers. 
Since the muon showers are correlated with the X-ray outbursts of the Cygnus 
X3, then it is logical to assume that hadronic particle packages are created 
in the same mechanism as outbursts or flares. Thus we can introduce the 
name of 'hadronic P Cygni' for stars with similar behavior and which are 
detected in the proper optical category.

So far our study has been based on regular physics of hadrons and quarks.
However, inside millisecond pulsars such extreme physical conditions may 
exist, when strong, electro-weak and gravitational interactions interplay 
with each other. It is challenging to describe the properties of these 
pulsars with unified models containing all of the general interactions. 
In this paper we introduce microscopical extra dimensions (Randall \& 
Sudrum 1999) and use the frame of a Kaluza\,--\,Klein like model, where the 
excited states of elementary particles appear as geometrical degrees 
of freedom (Kan \& Shiraishi 2000; Luk\'acs et al. 2003). Space like extra 
dimension offers a unique possibility to introduce strangeness 
as an excitation generated by the light quark moving into such a 
compactified extra dimension. We will investigate, if the properties of 
quark stars are modified in the presence of compactified extra space 
dimensions.

\section{Motion into the 5\textsuperscript{th} Dimension}

\begin{figure}[ht]
\centering
\hspace*{0.2truecm}\includegraphics[width=7.0cm]{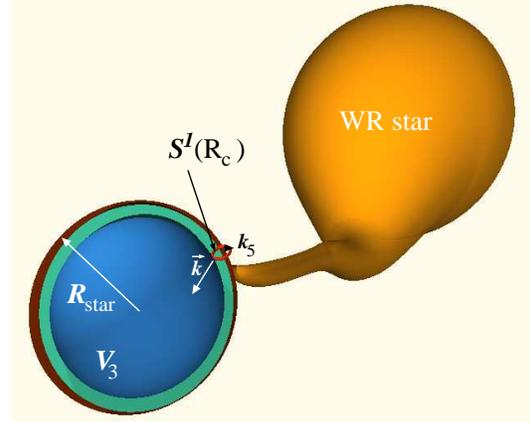}
\caption{Schematic view of the opening of the 5\textsuperscript{th} 
dimension on the Cygnus X3 binary system in a $3+1_c+1$ dimensional 
space-time, where the space has a 
$V_3 \times S^1$ structure with a compactified radius $R_c$ of the $S^1$ 
circle. The momentum for moving in the 5\textsuperscript{th} dimension 
is denoted by $k_5$\label{fig:1}.}
\end{figure}

Todays a new trend appeared in high energy physics, and the role of 
microscopical extra dimensions has been investigated from the point of view 
of high energy particle production. Although the symmetries of the 
observed 'low-energy' physical world suggest a space-time with $3+1$ 
macroscopical dimensions, we can not exclude the existence of 
compactified extra dimensions at microscopical scales. We can consider a 
$3+1_c+1$ dimensional space-time, where the particles have enough energy to 
move into the extra compactified space dimension indicated by $1_c$ 
(Barnaf\"oldi et al. 2003, 2004; Luk\'acs et al. 2003). 
Figure \ref{fig:1} illustrates the appearance of a compactified 
dimension.

Particles can move freely along the extra $x^5$ direction, however we 
require a periodic boundary condition, which results a Bohr-type 
quantization condition for $k_5$ momentum component. This condition induces 
an uncertainty in the position with the size of  $2 \pi R_c$, where $R_c$ 
is the compactification radius. Motion into the 5\textsuperscript{th} 
dimension generates an extra mass term appears in pure 4D description: 
\begin{equation}
k_5 = \frac{ n \,\,\hbar}{R_c} \,\,\,\,\,\,\,\ 
\longrightarrow \,\,\,\,\,\,\,\, 
\widehat{m}=\sqrt{m^2+\left(\frac{n \,\,\hbar}{R_c}\right)^2 }.
\label{quanta}
\end{equation}
Considering compactified radius $R_c \sim 10^{-12}- 10^{-13}$ cm this extra 
'mass' is ${\widehat m} \sim 100$ MeV, which is a familiar quantity in 
hadron spectroscopy (Arkhipov 2004).

One can introduce a 'pseudo charge' generated by $k_5$ with its  $\pm$ sign: 
\begin{equation}
        {\widehat q} = n \cdot \frac{ 2 \hbar \sqrt{G}}{c\,\,R_c} \ \ ,
\label{pseudoq}
\end{equation}
which acts in a vector-scalar interaction. We can directly see that 
${\widehat q}$ is {\it not} the electric charge. Indeed 
${\widehat q} \, ^2 < 16 \pi G m_0^2 $, where
$G$ is the gravitational constant and $m_0$ 
is the rest mass (Luk\'acs \& Pacher 1985).
This is either a familiar quantum number (e.g. strangeness, $S$), or 
another charge-like quanta, which is not yet observed
experimentally.

\section{Field Equations in 5D}

Our goal is to describe the inner structure of a {\it spherically  
symmetric} quark star-like object in $3+1_c+1$ D space-time. Our object 
is {\it static},  stays in equilibrium, and its temperature is close to 
{\it zero}. We are using a perfect fluid description with 
{\it isotropy} in the standard 3 spatial dimensions. 
{\it Killing-symmetries} are still exist in 5D space-time, but 
an {\it anisotropy} needs to be chosen for the fluid in the microscopical 
5\textsuperscript{th} dimension. These symmetries yield 
to the metric~\footnote{Notations: we 
are using coordinates $x^i$ with Latin indices ($i=0,1,2,3$ and $5$) for 
$3+1_c+1$ D space-time with the Einstein convention.}    
\begin{equation}\label{metric}
\dd s^2=
e^{2\nu} \, \, \dd t^2 - e^{2\lambda} \, \, 
\dd r^2 - r^2 \dd {\Omega}^2-e^{2\Phi }({\dd x^5})^2 \ ,
\end{equation}
where the quantities $\nu$, $\lambda$, and $\Phi$ depend only on the radius
$r$.  We denote the usual spherical elementary surface by $\dd \Omega^2$.
The 5D Einstein equation is written in the following way:
\begin{eqnarray}
-\gamma \ T_{ik} &=& R_{ik} - \frac{1}{2} R^l_{\,\,\, l} \ g_{ik}  \ \ .
\label{Einst}
\end{eqnarray}
Here the energy-momentum tensor contains 
the energy density $\varepsilon$, and the pressure components
$P$ and $P_5$ (but $P\neq P_5$):  
\begin{eqnarray}
T^{ik}  &=&
{\rm diag} \  ( \varepsilon \, e^{2 \nu},  \, P \, e^{2 \lambda},
 \, P\, r^2, \, P\,  r^2 \sin^2 \theta, \, P_5 \, e^{2 \Phi} ) 
\label{tik} \nonumber \\ 
\end{eqnarray}
Substituting eqs.~(\ref{metric}) and (\ref{tik}) into the 5D Einstein equation 
in eq.~(\ref{Einst}) one obtains the following equation system:
\begin{eqnarray}
&&{ -\gamma \ \varepsilon } = {- \frac{1}{r^2} } + \nonumber  \\
&& \ \ \ \   + {e^{-2 \lambda}} \left[  \Phi'' + \Phi'^2 - \lambda' \Phi'
  + \frac{2 \Phi'}{r} - \frac{2 \lambda'}{r} + \frac{1}{r^2} \right]
  \label{einst1}\,  \\
&&{ -\gamma \ P } = \frac{1}{r^2} +
  e^{-2 \lambda} \left[ { - \nu' \Phi'
  - \frac{2 \Phi'}{r}} - \frac{2 \nu'}{r} - \frac{1}{r^2} \right]
  \label{einst2} \\
&&{ -\gamma \ P } =
{ e^{-2 \lambda} \left[ -\nu'' -\nu'^2 + \nu' \lambda'
{ - \Phi'' - \Phi'^2 -} \right.} \nonumber \\
&& \ \ \ \ { \left. {
-\nu' \Phi' + \lambda' \Phi' 
- \frac{2 \Phi'}{r}}  - \frac{\nu'}{r} + \frac{2 \lambda'}{r} \right]
 }  \label{einst3} \,  \\
&& { -\gamma \ P_5 }  =  { \frac{1}{r^2} }+  \nonumber \\
&& \ \ \ \  + e^{-2 \lambda} \left[  -\nu'' -\nu'^2 + \nu' \lambda'
 - \frac{2\nu'}{r}  + \frac{2\lambda'}{r} - \frac{1}{r^2} \right]
  \label{einst4} \ \ 
\end{eqnarray}
where $\gamma = 8 \pi G / c^4$ and $G$ is the gravitational constant.

Assuming cold nuclear matter inside the fermion star ($T = 0$ 
approximation), all local material characteristics of the fluid
depend on one thermodynamic quantity, e.g. particle density, 
$\rho$. Thus, for the local matter we can write
\begin{equation}
\varepsilon = \varepsilon(\rho); \,\,\,\,\, \  P=P(\rho);\,\,\,\,\, \  P_5 = P_5(\rho) \ .
\end{equation}
{}From the appropriate Bianchi identity we obtain 
\begin{equation}
T^{ir}_{\,\,\,\,\,;r} = 0 \ \ \,\,\, \longrightarrow \ \ \,\,\, 
P' = -\nu '(\varepsilon + P) + (P_5-P) \Phi' \ \, . \, \label{bian}
\end{equation}
This equation clearly demonstrates the influence of the extra dimensional 
behavior on the normal isotropic pressure, where 
\mbox{$P=P_1=P_2=P_3$.}

In the 5D Einstein equations ~(\ref{einst1}-\ref{einst4})
two extra variables appeared, compared to the 4D formalism, namely 
$P_5$ and $\Phi$. However, $P_5(\rho)$ is a known
function of density and specified by the
actual interaction in the matter. Thus $\Phi(r)$ is the {\it only new degree 
of freedom} determined by eq.~(\ref{einst4}). Further details 
can be found in (Barnaf\"oldi, 2003).

\section{A Special 5D Solution}

We can recognize the similarity between 4D and 5D solutions.
For specially chosen pressure component $P_5$, there is a unique solution of 
the Einstein equations in eqs.~(\ref{einst1}-\ref{einst4}), namely 
$\dd \Phi / \dd r=0$. In this case eqs.~(\ref{einst1}-\ref{einst3}) 
lead to the Tolman\,--\,Oppenheimer\,--\,Volkov equation 
(Glendenning 1997) and 
can be solved separately with the $\Phi=$ const. condition. The  
eq.~(\ref{einst4}) gives $P_5$. 
Although these solutions do not differ formally from the 
4D neutron star solution (except for $P_5$),
but the extra dimension has its influence
on $\varepsilon(\rho)$ and $P(\rho)$ (Kan \& Shiraishi 2000).
In case of $\Phi \neq $ const. condition, 
one must turn to numerics and solve directly
eqs. (\ref{einst1}-\ref{einst4}). 

Neglecting the effects of electromagnetic charge, let us start  
with a neutral, single massive fermion ('neutron', $N$) as 
elementary building block of a compact star.
Since the minimal nonzero fifth momentum component is given by 
Bohr-type quantization~(\ref{quanta}), then the extra direction 
of the phase space is not populated until the Fermi-momentum 
$k_F < \hbar/R_c$. However, at the threshold both $k_5 = \pm \hbar /R_c$ 
states appear. One can represent this as another ('excited') particle 
with mass $m_X = \sqrt{m_N^2 + (\hbar/R_c )^2 }$ (with a non-electric 
'charge' ${\widehat q} = \pm \frac{2 \hbar \sqrt{G}}{c R_c}$ as well).
The equations obtain a form as if this second particle appears in
complete chemical equilibrium with the neutron: $\mu_X = \mu_N$.

This recipe is repeated for every integer $n$, when $k$ exceeds 
a threshold $n\hbar / R_c$, and these higher excitations are introduced into
the equation of state.
\vspace*{-0.6truecm}
\begin{figure}[ht]
\centering
\includegraphics[width=7.5cm]{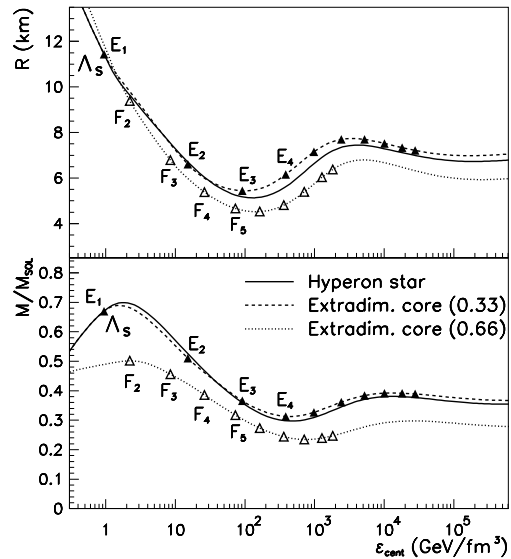}
\caption{$M(\varepsilon_{cent})$ and $R(\varepsilon_{cent})$ functions
for 4D hyperon star (solid lines) and 5D hybrid stars 
(dashed and dotted lines).}
\label{fig:2}
\end{figure}

We apply our model for hybrid star with strangeness content as a reference.
This object is a hyperon star in 4D with $N$ and $\Lambda_s$ content.
After solving eqs.~(\ref{einst1})-(\ref{einst3}),
Fig. \ref{fig:2} displays the central energy density 
($\varepsilon_{cent}$) dependence of
the star mass and star radius in this scenario (solid lines).

Next we introduce an extra compactified space dimension ($1_c$)
with radius $R_c=0.33$ fm. In this case the particle
$\Lambda_s$ can be interpreted as the first extra dimensional
excited state of $N$. Solving eqs.~(\ref{einst1})-(\ref{einst4}) 
we obtain similar results to our reference hyperon star (dashed lines).
The location of the appearances of extra dimensional excitations
are denoted by the letters $E_1 (\equiv \Lambda_s), \ E_2, \ E_3, \ \dots$
on Fig. \ref{fig:2}.

Finally we vary the compactification radius and increase it to
$R_c=0.66$ fm. Solving the TOV-like 
equation system,
we obtain slightly different curves for the star mass and radius (dotted lines),
however the main characteristics remain the same.
In Fig. \ref{fig:2} these slightly lighter extra-dimensional
excitations are denoted by the 
letters $F_1$,  $F_2$,  $F_3, \ \dots$.

\begin{figure}[ht]
\centering
\includegraphics[width=7.5cm]{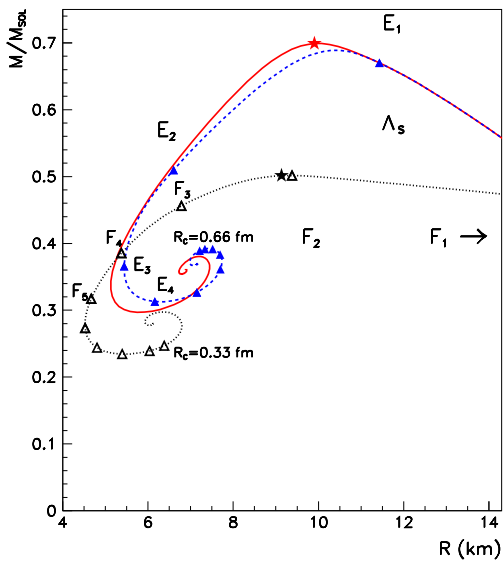}
\caption{The $M(R)$ diagrams for 4D hyperon star and 5D hybrid stars.
See text for details.}
\label{fig:3}
\end{figure}

We construct the $M(R)$ diagrams for all three scenarios and display
them on Fig. \ref{fig:3}. The introduction of a proper extra dimensional
radius ($R_c=0.33$ fm) yields to a similar object to our reference
hyperon star. This figure shows, that the change 
in the compactification radius is able to modify the
$M(R)$ diagram. The study of the limits and consequences
of these modifications are under investigation.

\vspace*{-0.6truecm}

\vspace*{-0.2truecm}
\section{Discussion}

Introducing extra dimensions into the description of hybrid stars
led to similar results in $3+1$ and $3+1_c+1$ space-time.
The $M(R)$ diagram does not change drastically, if the extra
dimensional radius is in the same order as the strange quark
mass difference ($\Delta m_s \sim \hbar c/R_c$). However, the
presence of one (or more) compactified dimension(s) may able to change the 
characteristics of these astrophysical objects.
We are aware of the complexity of this question, 
and do not want to claim the extra dimensional origin
of the strangeness. However such an astrophysical object 
as a compact or hybrid star serves as 'natural laboratory' 
to investigate these problems (Weber 1999)
or other exotic ones, e.g. the flavor dependence of
gravity (Fischbach et al. 1986, 1991).
In the Earth the planned experiments at the Large Hadronic Collider (LHC)
may shed of some light on this question in the near future.
We are aware of the consideration of the rotation for
the investigated astrophysical objects, but the correct calculation
and the necessary statibility investigations require a very complex
treatment, which is beyond the scope of our recent paper.

\acknowledgements
This work was supported by the E\"otv\"os University 
Department of Astronomy, 
British Council and 
Hungarian OTKA grants {T047050} and  {NK62044}. 



\begin{thebibliography}{00}

\bibitem{00}
Allison et al.: 1999, Proc. of the 26\textsuperscript{th} Int. Cosmic Ray 
Conference (ICRC 99), Salt Lake City, Utah (arXiv:hep-ex/9905045) 

\bibitem{01}
Ambartsumyan, V.~A., Saakyan, G.~S.: 1960,  Astron. Zh. 37, 193

\bibitem{02}
Arkhipov, A.~A.: 2004, AIP Conf. Proc. 717, 680

\bibitem{03}
Aronson, S.H. et al.: 1986, Phys. Rev. Lett. 56, 1342 

\bibitem{04}
Barnaf\"oldi, G.~G., L\'evai, P.,  Luk\'acs, B.: 2003,  
in Proceedings of the 4th Int. Workshop on New Worlds in
      Astroparticle Physics, Faro, Portugal,
Worlds Scientific, Singapoure;

\bibitem{04a}
Barnaf\"oldi, G.~G., L\'evai, P.,  Luk\'acs, B.: 2004,  
Publ. of the Astronomy Department of Astronomy
of the E\"otv\"os University, PADEU 14, Budapest, Hungary

\bibitem{05}
Baym, G. et al.: 1985, Phys. Lett. 160B, 181

\bibitem{06}
BNL --- Brookhaven National Laboratory: 2000, 
See WEB-page http://www.phy.bnl.gov/newphysics/experiments.html

\bibitem{06a}
Collins A.: 2000, See WEB-page http://www.andrewcollins.com 

\bibitem{07}
Fischbach, E. et al.: 1986, Phys. Rev. Lett. 56, 3 

\bibitem{07a}
Fischbach, E. et al.: 1996, Les Arcs 1996, Dark matter in cosmology, 
quantum measurements, experimental gravitation,  443 

\bibitem{08}
Glendenning, N.~K.: 1997, Compact Stars, Springer, NH

\bibitem{09}
Grassi, F.: 1988, FERMILAB-Pub-88/19-A


\bibitem{11}
Hillas, A.~M.: 1984, Ann. Rev. Astrophys, 22, 425

\bibitem{12}
Kan, N., Shiraishi, K.: 2002,  Phys. Rev. D66, 105014 

\bibitem{13}
Kondratiuk, L.~A., Krivoruchenko, M.~I., Shchepkin, M.~G.:
1986, JETP Letters (ISSN 0021-3640), 43, 10 

\bibitem{14}
Luk\'acs, B., Barnaf\"oldi, G.~G., L\'evai, P.: 2003  
in Proc. of NATO ARW on Superdense QCD Matter
 and Compact Stars, Yerevan, Armenia,  
NATO ASI Series 197, Blascke, D., Sedrakian, D. (eds.), Kluwer

\bibitem{15}
Luk\'acs, B.: 2000, Relativity Today, C. Hoenselaers, Z. Perj\'es (eds.), 
Akad\'emiai Kiad\'o, Budapest, 161

\bibitem{16}
Luk\'acs, B., Pacher, T.: 1985,  Phys. Lett. A113, 200 


\bibitem{17}
Marshak, M.~L. et al: 1985, Phys. Rev. Lett. 54, 2079 {\it ibid} 55, 1965  

\bibitem{18}
Marshak, M.~L. et al.: 2000,  
Proc. of the International Conference of High Energy Physics
(ICHEP 2000), see http://ichep2000.hep.sci.osaka-u.ac.jp/abs$\_$PA-11.html

\bibitem{18a}
Mart\'\i, J., Paredes, J.~M., Peracaula, M.: 2001, 
Astron. \& Astrophys. 375, 476 

\bibitem{18b}
Miller-Jones et al.: 2004, Astrophys. J. 600, 368

\bibitem{19}
Olinto, A.~V.: 1991, FERMILAB-Conf-91/202-A; 349-A

\bibitem{20}
Priszny\'ak, M. Luk\'acs B., L\'evai, P.: 1994, 
Report KFKI-1994-24/A, KFKI, Budapest 


\bibitem{21}
Randall, L., Sudrum, R.: 1999, Phys. Rev. Lett. 83, 3370 {\it ibid} 4690

\bibitem{22}
Weber, F.: 1999, Astrophysical Laboratories for Nuclear and Particle 
Physics, IOP Publishing, Bristol, England

\bibitem{23}
Witten, E.: 1984, Phys. Rev. D30, 272  

\end{thebibliography}
\end{document}